\pdfoutput=1

\documentclass[11pt,twoside,a4paper,cmspaper,final,collab]{cms-tdr}
\def\svnVersion{79668:79669}\def\cmsCernNo{2011-137}\def\cmsCernDate{\today}\def\cmsMessage{Submitted to the Journal of Instrumentation}

\begin{document}\cmsNoteHeader{TAU-11-001}

\hyphenation{had-ron-i-za-tion}
\hyphenation{cal-or-i-me-ter}
\hyphenation{de-vices}
\RCS$Revision: 79669 $
\RCS$HeadURL: svn+ssh://alverson@svn.cern.ch/reps/tdr2/papers/TAU-11-001/trunk/TAU-11-001.tex $
\RCS$Id: TAU-11-001.tex 79669 2011-09-27 20:30:07Z alverson $

\newcommand{\METx}{\ensuremath{E^{\mathrm{miss}}_{x}}\xspace}
\newcommand{\METy}{\ensuremath{E^{\mathrm{miss}}_{y}}\xspace}
\newcommand{\PZ}    {\mathrm{Z}}

\newcommand{\tablesize}{\small}

\newcommand{\captiontext}{}

\cmsNoteHeader{TAU-11-001} 
\title{Performance of \texorpdfstring{$\tau$}{tau}-lepton reconstruction and identification in CMS} 

\address[umwis]{
  Department of Physics,
  University of Wisconsin,
  Madison, WI 53706-1390
}

\address[LLR]{
Laboratoire Leprince-Ringuet, \'Ecole Polytechnique,
91128 Palaiseau,
France 
}

\address[saclay]{
DSM/IRFU/SPP, CEA Saclay, 
91191 Gif-sur-Yvette, France
}

\address[rwthaachen]{
  1. Physikalisches Institut B,
  RWTH Aachen University,
  D-52074 Aachen
}

\address[ucdavis]{
  Department of Physics,
  University of California,
  Davis, CA 95616--8677
}

\address[IPJ]{
 The Andrzej Soltan Institute for Nuclear Studies, 05-400 Swierk/Otwock, Poland 
}

\address[cern]{
  European Organization for Nuclear Research, 
  Geneva, 
  Switzerland
}

\address[ipm_and_shiraz]{
  School of Particles and Accelerators, 
  Institute for Studies in Theoretical Physics and Mathematics (IPM),
  Tehran 19395, 
  Iran; 
  also at: \\
  Physics Department, 
  Shiraz University, 
  Shiraz 71454, 
  Iran 
}

\address[lisbon]{
  LIP--Laboratory for Instrumentation and Experimental Particle Physics,
  Lisbon, 
  Portugal
}

\author[umwis]{M.~Bachtis}
\author[LLR]{L.~Bianchini}
\author[saclay]{S.~Choudhury}
\author[umwis]{S.~Dasu}
\author[rwthaachen]{M.~Edelhoff}
\author[ucdavis]{E.~K.~Friis}
\author[IPJ]{T.~Fruboes}
\author[cern]{S.~Gennai}
\author[ucdavis]{S.~Maruyama}
\author[ipm_and_shiraz]{A.~Mohammadi}
\author[lisbon]{A.~Nayak}
\author[umwis]{A.~Savin}
\author[umwis]{J.~Swanson}
\author[ucdavis]{C.~Veelken}

\date{\today}

\abstract{
The performance of $\tau$-lepton reconstruction and identification algorithms is
studied using a data sample of proton-proton collisions at $\sqrt{s}=7$~TeV,
corresponding to an integrated
luminosity of 36 pb$^{-1}$ collected with the
CMS detector at the LHC. The $\tau$ leptons that decay into one or three charged
hadrons, zero or more short-lived neutral hadrons,
and a
neutrino are identified using final-state particles reconstructed in
the CMS tracker and electromagnetic calorimeter.  The
reconstruction efficiency of the algorithms is measured using
$\tau$ leptons
produced in Z-boson decays. The
$\tau$-lepton misidentification rates for jets and electrons are determined.
}

\hypersetup{%
pdfauthor={CMS Collaboration},%
pdftitle={Performance of tau-lepton reconstruction and identification in CMS},%
pdfsubject={CMS},%
pdfkeywords={CMS, physics, software, computing}}

\maketitle 
\def\verPAPER{1}
\def\verAN{2}
\def\ver{1}

\ifx\ver\verPAPER

\section{Introduction}

The primary goal of the Compact Muon Solenoid (CMS)~\cite{CMSExperiment} experiment
is to
explore particle
physics at the TeV energy scale by studying the final states produced in
the proton-proton collisions
at the Large
Hadron Collider (LHC)~\cite{LHC}.
Leptons play a very important role
in these studies because they often represent
an experimentally favourable signature.

The three generations of
 charged leptons, electrons, muons, and taus, are characterized by
their masses.
Because of their higher mass, $\tau$ leptons play a crucial
role in the searches for the standard model (SM) Higgs boson, especially for
the  mass region below
 twice the $\PW$-boson mass.  The motivation for
searches for the Higgs boson in its $\tau$-leptonic decays is also
supported for example by the minimal supersymmetric standard model (MSSM)~\cite{Martin97}.
Other models of new physics, such as sypersymmetric left-right models (SUSYLR),
 also predict increased couplings to the
third-generation charged fermions. As a result, the decay chains of the
supersymmetric particles lead to the lighter stau, which can lead to multi-tau
final states~\cite{SUSYLR}.
Lepton universality ensures that
one third of $\PW$ and $\PZ$-boson leptonic decays result in
$\tau$ leptons. When measuring rare processes, this contribution
becomes substantial. For example, in the search for high-mass
SM Higgs bosons
 that decay preferentially into $\PW$ and $\PZ$ bosons,
the addition of modes with $\tau$ leptons in the final state improves the
early discovery potential.

The lifetime of $\tau$ leptons is short enough that they decay before
reaching the detector elements.
In two thirds of the cases, $\tau$ leptons decay
hadronically, typically into one or three charged mesons
(predominantly $\pi^{+}$, $\pi^{-}$), often accompanied by neutral pions
(decaying via $\pi^{0}~\to~\gamma\gamma$),
 and a
$\nu_\tau$.

The CMS collaboration
has designed algorithms that
use final-state photons and charged had\-rons to
identify
hadronic decays of $\tau$ leptons ($\tau_{\rm{h}}$)
through the reconstruction
of the intermediate resonances.
The $\nu_\tau$
escapes undetected and is not considered in the $\tau_{\rm{h}}$ reconstruction.
These algorithms use
 decay mode identification techniques and efficiently discriminate against
 potentially large backgrounds from quarks and gluons that occasionally hadronize into jets of  low
 particle multiplicity.
 The algorithms described here
have already been successfully used in a measurement of the $\PZ\to\tau\tau$
production cross section~\cite{ourZtautaupaper} and
in a search for neutral MSSM Higgs bosons decaying into
$\tau$ pairs~\cite{ourHtautaupaper}.

This paper describes
performance studies
based on a sample of proton-proton collisions collected during 2010
at $\sqrt{s}=7$~TeV,
corresponding to an integrated luminosity of
36~pb$^{-1}$.
The analysis uses genuine taus from inclusive $\PZ \to \tau\tau$ production.
One tau is required to decay leptonically, into a muon,
and the other one hadronically, thus creating a $\mu\tau_{\rm{h}}$ final state.
The analysis provides estimates of the $\tau_{\rm{h}}$ reconstruction and
identification efficiency, and determines the misidentification rate, the
probability for quark and gluon jets or electrons to be misidentified as
$\tau_{\rm{h}}$.
This paper uses the selection requirements that are most commonly used in the $\PZ$ and Higgs
analyses, and compares the LHC collision data with predictions based on
Monte Carlo (MC) simulation.

\section{CMS Detector}

A detailed description of CMS can be found elsewhere~\cite{CMSExperiment}.
The central feature of the CMS apparatus
is a superconducting solenoid of 6~m internal diameter, providing
a magnetic field of 3.8~T. Within the field volume are the silicon pixel
and strip tracker, the crystal electromagnetic calorimeter (ECAL),
and the brass/scintil\-lator hadron calorimeter.
Muons are measured in gas-ionization detectors embedded
in the steel return yoke.

CMS uses a right-handed coordinate system, with the
origin at the nominal interaction point, the $x$ axis
pointing to the centre of the LHC ring, the $y$ axis pointing
up perpendicular to the LHC plane, and the $z$ axis
along the counterclockwise beam direction. The polar angle
$\theta$ is measured from the positive $z$ axis and the
azimuthal angle $\phi$ is measured in the $x$-$y$ plane.
Variables used in this article are
the pseudorapidity,
$\eta \equiv -\ln[\tan(\theta/2)]$, and the transverse
momentum, $\PT = \sqrt{p_x^2 + p_y^2}$.

The ECAL is designed to have both excellent energy resolution
and high granularity, properties that are crucial for reconstructing
electrons and photons produced in $\tau$-lepton decays.
The ECAL is constructed with projective lead tungstate
crystals in two pseudorapidity
regions: the barrel ($|\eta| < 1.479$) and the endcap
($1.479 < |\eta| < 3$).
In the barrel region, the crystals
are $25.8X_0$ long, where $X_0$ is the radiation length, and provide a granularity of
$\Delta\eta\times\Delta\phi = 0.0174\times0.0174$.
The endcap region is instrumented with a lead/silicon-strip
preshower detector consisting of two orthogonal strip
detectors with a strip pitch of 1.9~mm.  One plane is
at a depth of $2X_0$ and the other at $3X_0$.
The ECAL has an energy resolution of better than $0.5\%$ for unconverted photons with transverse energies above 100\GeV.

The inner tracker measures charged particle tracks
within the range $|\eta| < 2.5$.  It consists of
$1\, 440$ silicon pixel and $15\, 148$ silicon strip detector modules,
and provides an impact parameter resolution of
$\sim$\,15~$\Pgm$m and a transverse momentum
resolution of about $1.5\%$ for
100\GeV particles.
The reconstructed tracks are used to measure the location of interaction vertex(es).
The spatial resolution of the reconstruction is
$\approx 25 \mu m$ for vertexes with more than 30 associated tracks~\cite{tracker}.

The muon barrel region is covered
by drift tubes, and the endcap regions by cathode
strip chambers.  In both regions, resistive plate
chambers provide additional coordinate and timing
information.  Muons can be reconstructed in the
range $|\eta| < 2.4$, with a typical $\PT$
resolution of $1\%$ for $\PT \approx 40$\GeVc.

\section{CMS \texorpdfstring{$\tau_{\rm{h}}$}{tau(h)} Reconstruction Algorithms}
\label{algorithms}

CMS has developed two algorithms for identifying $\tau_{\rm{h}}$ decays,
based on the categorization of the $\tau_{\rm{h}}$-decay
 channels through the reconstruction of intermediate resonances:
the hadron plus strips (HPS) and
 the tau neural
classifier (TaNC) algorithms.
The HPS algorithm is used as the main algorithm in most previous
CMS $\tau$ analyses, with TaNC used for crosschecks.
Both
algorithms use particle flow (PF~\cite{PAS_PFT-09-001}) particles.  In the
PF approach, information from all subdetectors is combined
to reconstruct and identify all particles produced in the collision.
The particles are classified into mutually exclusive
categories:
charged hadrons, photons, neutral hadrons, muons, and
electrons.
These algorithms
are designed to optimize the performance of the
$\tau_{\rm{h}}$ identification and reconstruction
by considering the different
hadronic decay modes of the tau individually.
The dominant hadronic decays of
$\tau$ leptons consist of one or three charged $\pi$ mesons and up to two $\pi^0$
mesons, as summarized in Table~\ref{tab:decay_modes}.

\begin{table}[htbp]
\begin{center}
\caption{\captiontext
   Branching fractions of the dominant hadronic decays of
   the $\tau$ lepton and the symbol and mass of any intermediate resonance~\cite{PDG}.
   The $h$ stands for
   both $\pi$ and $K$, but in this analysis the $\pi$ mass is assigned to all
   charged particles. The table is symmetric under charge conjugation.}
\tablesize
\begin{tabular}{|l|c|c|c|}
\hline
Decay mode & Resonance & Mass (\!\MeVcc) &  Branching fraction (\%) \\
\hline
$\tau^{-}$  $\rightarrow $  $h^{-} \nu_{\tau}$ &  &  & $11.6\%$ \\
$\tau^{-}$  $\rightarrow $  $h^{-} \pi^{0}  \nu_{\tau}$ & $\rho^{-}$ & 770 & $26.0\%$ \\
$\tau^{-}$  $\rightarrow $  $h^{-} \pi^{0}\pi^{0}  \nu_{\tau}$ & $a_{\rm{1}}^{-}$ & 1200 & $9.5\%$ \\
$\tau^{-}$  $\rightarrow $  $h^{-} h^{+} h^{-} \nu_{\tau}$ & $a_{\rm{1}}^{-}$  & 1200 & $9.8\%$ \\
$\tau^{-}$  $\rightarrow $  $h^{-} h^{+} h^{-}\pi^{0}  \nu_{\tau}$ & & & $4.8\%$ \\
      \hline
\end{tabular}
\label{tab:decay_modes}
\end{center}
\end{table}

Both algorithms start the reconstruction of a $\tau_{\rm{h}}$ candidate from a PF jet, whose
four-momentum is reconstructed using the anti-$k_{\rm{T}}$
algorithm with a distance parameter $R = 0.5$~\cite{AntiKT}.
Using a PF jet as an initial seed, the algorithms first reconstruct the $\pi^0$ components
of the $\tau_{\rm{h}}$, then combine them with charged hadrons to reconstruct the
tau decay mode and calculate the tau four-momentum and isolation quantities.

\subsection{HPS Algorithm}
\label{HPS_description}

The HPS algorithm gives
special attention to photon conversions in the CMS tracker material.
The bending of electron/positron tracks
in the magnetic field of the CMS solenoid
broadens the calorimeter signatures of neutral pions in
the azimuthal direction.
This effect
is taken into account in the HPS algorithm by reconstructing photons in ``strips'',
objects that are built out of electromagnetic particles (PF photons and electrons).
The strip reconstruction starts by centering a strip
on the most energetic electromagnetic particle
within
the PF jet.
The algorithm then searches for other electromagnetic particles
within a window of size $\Delta \eta = 0.05$ and $\Delta \phi = 0.20$
centered on the strip center.
If other electromagnetic particles are found within that window,
the most energetic one gets associated with the strip
and the strip four-momentum is recalculated.
The procedure is repeated
until no further particles are found that can be associated with the strip.
Strips satisfying a minimum transverse momentum requirement of $\pt^{\rm{strip}} > 1$\GeVc
are finally combined with the charged hadrons to reconstruct individual
$\tau_{\rm{h}}$ decay modes.

The decay topologies that are considered by the HPS tau identification algorithm are
\begin{enumerate}
\item{\it Single hadron}
      corresponds to $ h^{-} \nu_{\tau}$
      and $ h^{-} \pi^{0} \nu_{\tau}$ decays
      in which the neutral pions have too little energy to be reconstructed as strips.
\item{\it One hadron $+$ one strip}
      reconstructs the decay mode $ h^{-} \pi^{0} \nu_{\tau}$
      in events in which the photons from $\pi^{0}$ decay
      are close together on the calorimeter surface.
\item{\it One hadron $+$ two strips}
      corresponds to the decay mode $ h^{-} \pi^{0} \nu_{\tau}$
      in events in which photons from $\pi^{0}$ decays are well separated.
\item{\it Three hadrons}
      corresponds to the decay mode $ h^{-} h^{+} h^{-} \nu_{\tau}$.
      The three charged hadrons are required
      to come from the same secondary vertex.
\end {enumerate}

There are no separate decay topologies for
the  $h^{-} \pi^{0} \pi^{0}$ and $ h^{-} h^{+} h^{-} \pi^{0} \nu_{\tau}$ decay modes.
They are reconstructed via the existing topologies.
All charged hadrons and strips are required to be contained within a cone of size
$\Delta R = (2.8\GeVc) /\pt^{\tau_{\rm{h}}}$,
where $\pt^{\tau_{\rm{h}}}$ is the transverse momentum of the reconstructed $\tau_{\rm{h}}$.
The reconstructed tau momentum $\vec{p}^{\tau_{\rm{h}}}$
is required to match the ($\eta$, $\phi$) direction
of the original PF jet within a maximum distance of $\Delta R = 0.1$,
where $\Delta R = \sqrt{(\Delta\eta)^2+(\Delta\phi)^2}$.

The four-momenta of charged hadrons and strips are reconstructed according to the respective
$\tau_{\rm{h}}$
decay topology hypothesis, assuming all charged hadrons to be pions, and
are required to be consistent with the masses of the intermediate meson
resonances listed in Table~\ref{tab:decay_modes}.
The following invariant mass windows are allowed for candidates:
50 -- 200\MeVcc
for $\pi^0$,
0.3 -- 1.3\GeVcc
 for $\rho$, and
0.8 -- 1.5\GeVcc for $a_{\rm{1}}$.
In cases where a $\tau_{\rm{h}}$ decay is consistent with more than
one hypothesis,
the hypothesis giving the highest $\pt^{\tau_{\rm{h}}}$
is chosen.

Finally, reconstructed candidates are required to be isolated.
The isolation criterion
requires that,
apart from the
$\tau_{\rm{h}}$ decay products, there be
no
charged
hadrons or photons present
within an isolation cone of size $\Delta R =
0.5$ around the direction of the $\tau_{\rm{h}}$.
By adjusting the $\pt$ threshold for particles that are considered
 in the isolation cone,
three
working points, "loose", "medium", and "tight" are defined.
The working points are determined using a simulated sample of QCD dijet events.
The ``loose'' working point corresponds to a probability of approximately 1\%
for jets to be misidentified as $\tau_{\rm{h}}$. Successive working
points reduce the misidentification rate by a factor of two with respect to the previous one.

\subsection{TaNC Algorithm}
\label{TaNC_description}

In the TaNC case
the leading (highest-\pt) particle is required to have a \pt above  $5$\GeVc
and
to be within  $\Delta R = 0.1$
around the jet direction.
The PF $\tau_{\rm{h}}$ four-momentum is reconstructed as a
sum of the four-momenta of all particles with \pt above $0.5$\GeVc in a
cone of radius $\Delta R = 0.15$ around the direction of the leading
particle.
A signal cone size is defined to be $\Delta R_{\rm photons} = 0.15$ for photons and
$\Delta R_{\rm charged} = (5\GeV)/E_{\rm T}$ for charged hadrons, where $E_{\rm T}$
is the transverse energy of the PF $\tau_{\rm{h}}$, and
$\Delta R_{\rm charged}$ is restricted to be within the range
$ 0.07 \leq \Delta R_{\rm charged} \leq 0.15$.
The signal cone is the region where the $\tau_{\rm{h}}$
decay products are expected to be found.
An isolation annulus is defined between the signal cone and a wider
isolation cone
of outer radius $\Delta R = 0.5$ around the leading particle.

The decay mode is reconstructed from the particles that are contained within the
signal cone of the $\tau_{\rm{h}}$ candidate by counting the number of tracks and $\pi^0$
meson candidates.  The $\pi^0$ meson candidates are reconstructed
by merging pairs of photons that have an invariant mass of less than
$0.2$\GeVcc.  All unpaired photons are considered as $\pi^0$ candidates
if their \pt exceeds 10\% of
the PF $\tau_{\rm{h}}$ transverse momentum.

The decay mode of each $\tau_{\rm{h}}$ candidate is uniquely determined by the multiplicity
of reconstructed objects in the signal cone.  Candidates with decay topologies
 other than those listed in
Table~\ref{tab:decay_modes} are immediately rejected.  Otherwise, a neural network is
used to compute a discriminant quantity for the $\tau_{\rm{h}}$ candidate.  Each decay
mode of Table~\ref{tab:decay_modes} uses a different neural network.  The input
observables used for each neural network are optimized for the topology of the decay
mode, and are constructed from the four-momenta of the particles in the  signal cone and the isolation annulus.
 In general, the signal cone input observables are chosen to parameterize the decay kinematics
of the intermediate resonance, and the isolation cone observables to describe the
multiplicity and \pt spectrum of nearby particles.
The variables include angular correlations between different particles within the signal and the isolation
cones, invariant masses calculated using different combinations of the particles, transverse momenta,
and numbers of charged particles in the signal and the isolation regions.
The neural networks are trained to discriminate between
genuine $\tau_{\rm{h}}$  produced in  $\PZ \to \tau \tau $ decays
 and misidentified jets from a sample of QCD
multijet events.
The set of input observables
for a given neural network is chosen to be the minimal set of observables for
which the removal of any two input variables significantly degrades the
classification performance.

The output of the neural network is a continuous quantity.
By adjusting the thresholds of selections on the neural network output, three
working points, again called ``loose'', ``medium'', and ``tight'', are defined,
 similar to those discussed in Section~\ref{HPS_description}.

\section{Efficiency of \texorpdfstring{$\tau_{\rm{h}}$}{tau(h)} Reconstruction and Identification}

To compare the performance of $\tau_{\rm{h}}$ reconstruction in data and MC simulation,
a set of MC samples is used to reproduce a mixture of signal and
background events. The signal is expected to come from inclusive $\PZ \to \tau\tau$
production. The major sources of background are $\tau\tau$ Drell--Yan production outside
of the $\PZ$-mass region, $\PW$ production with associated jets,
QCD multijet, and $t\bar{t}$ production.
The
Drell--Yan signal and background are simulated with the next-to-leading order (NLO)
MC generator
{\sc POWHEG}~\cite{Alioli:2008gx,Nason:2004rx,Frixione:2007vw}.
The QCD multijet and  $\PW$
backgrounds are simulated
with {\sc PYTHIA}~\cite{Sjostrand:2006za}
and the top quark samples with Madgraph~\cite{Maltoni:2002qb}.
The $\tau$-lepton decays are simulated with Tauola~\cite{Jadach:1993hs}.
The samples are normalized using the cross section
at next-to-next-to-leading order (NNLO) for
Drell--Yan and $\PW$, at leading order (LO) for QCD,
and NLO for the $t\bar{t}$ sample.
The MC samples are  mixed
based on
the corresponding cross sections.

To measure the efficiency of $\tau_{\rm{h}}$ reconstruction and identification in
data, a tag-and-probe method is used with
a sample of $\PZ\to\tau\tau\to\mu\tau_{\rm{h}}$ events.
The events are preselected using kinematic cuts and a set of
requirements to suppress the background from $\PZ \rightarrow \mu \mu$,
$\PW$, and QCD events, but without applying the $\tau_{\rm{h}}$-identification
algorithms.  The preselection requires the event to be triggered by a
single-muon high level trigger~\cite{HLT}, and to contain only one isolated muon
with $\pt^{\mu}
> 15$\GeVc within the geometric acceptance $\left| \eta_{\mu} \right|
< 2.1$,
that is used as a tag.
An isolated jet candidate of $\pt^{\rm{jet}} > 20$\GeVc
within the geometric acceptance $\left| \eta_{\rm{jet}} \right| < 2.3$,
with a ``leading'' (highest-$\pt$) track constituent in the jet with $\pt >
5$\GeVc, is used as a probe.
The preselection is needed to increase the percentage of $\PZ \rightarrow \tau \tau$ events in the final sample. This preselection clearly biases the
sample, but the bias is taken into account when computing the final efficiency.
The four-momentum of the jet is reconstructed using the anti-$k_{\rm{T}}$
algorithm with a
distance parameter of  0.5~\cite{AntiKT}.
The muon and the ``leading'' track in the
jet are required to be of opposite charge.  To suppress background
from $\PW$+jet(s) events,
an additional requirement on transverse mass, $M_{\rm{T}}$, of the muon
and missing transverse energy, $\ET^{\rm{miss}}$, of less than
40\GeV is applied. The transverse mass is defined as $M_{\rm{T}} = \sqrt{
  2 \PT^{\mu} \ET^{\rm{miss}} \cdot \left ( 1 - \cos {\Delta \phi} \right )
}$, where $\PT^{\mu}$ is the muon transverse momentum
and $ \Delta \phi$ is the azimuthal
angle between the $\ET^{\rm{miss}}$ vector and $\PT^{\mu}$.

The HPS and TaNC algorithms are both applied to the preselected events. The
resulting invariant mass distributions of the $\mu$-jet
system for those events that pass or fail the $\tau_{\rm{h}}$ identification
are fitted using signal and background distributions provided by MC
 simulation.  The efficiency is then calculated as
$\varepsilon =
  {N^{Z \to \tau \tau}_{\rm{pass}}}/{(N^{Z \to \tau \tau}_{\rm{pass}} + N^{Z \to \tau \tau}_{\rm{fail}})}$,
where $N^{Z \to \tau \tau}_{\rm{pass,fail}}$ are the numbers of $\PZ\to\tau\tau$ events after background contributions are subtracted.
Figure~\ref{fig:efficiency} shows
 the invariant mass of the $\mu$-jet system
for preselected events that pass (left) and fail (right)  the ``loose'' $\tau_{\rm{h}}$ identification requirements.
Since in the ``failed'' sample
there is no $\tau_{\rm{h}}$ reconstructed,
for consistency
the visible mass is always computed using the jet four-vector
and not the four-vector as reconstructed by the $\tau_{\rm{h}}$ algorithms.
The MC predictions
for signal and background events are also shown.
The ``passed'' sample is
dominated by $\PZ$ events and a small background contribution.
The sample of ``failed'' events is dominated by background contributions.
The MC predictions describe the data reasonably well.
The stability of the fit results is tested by using background estimates from data
instead of the MC
predictions and by varying the invariant mass ranges for the fit. All checks demonstrate
 consistent results
within the uncertainties of the method.

\begin{figure}[hbtp]
  \begin{center}
  \includegraphics[width=0.48\textwidth]{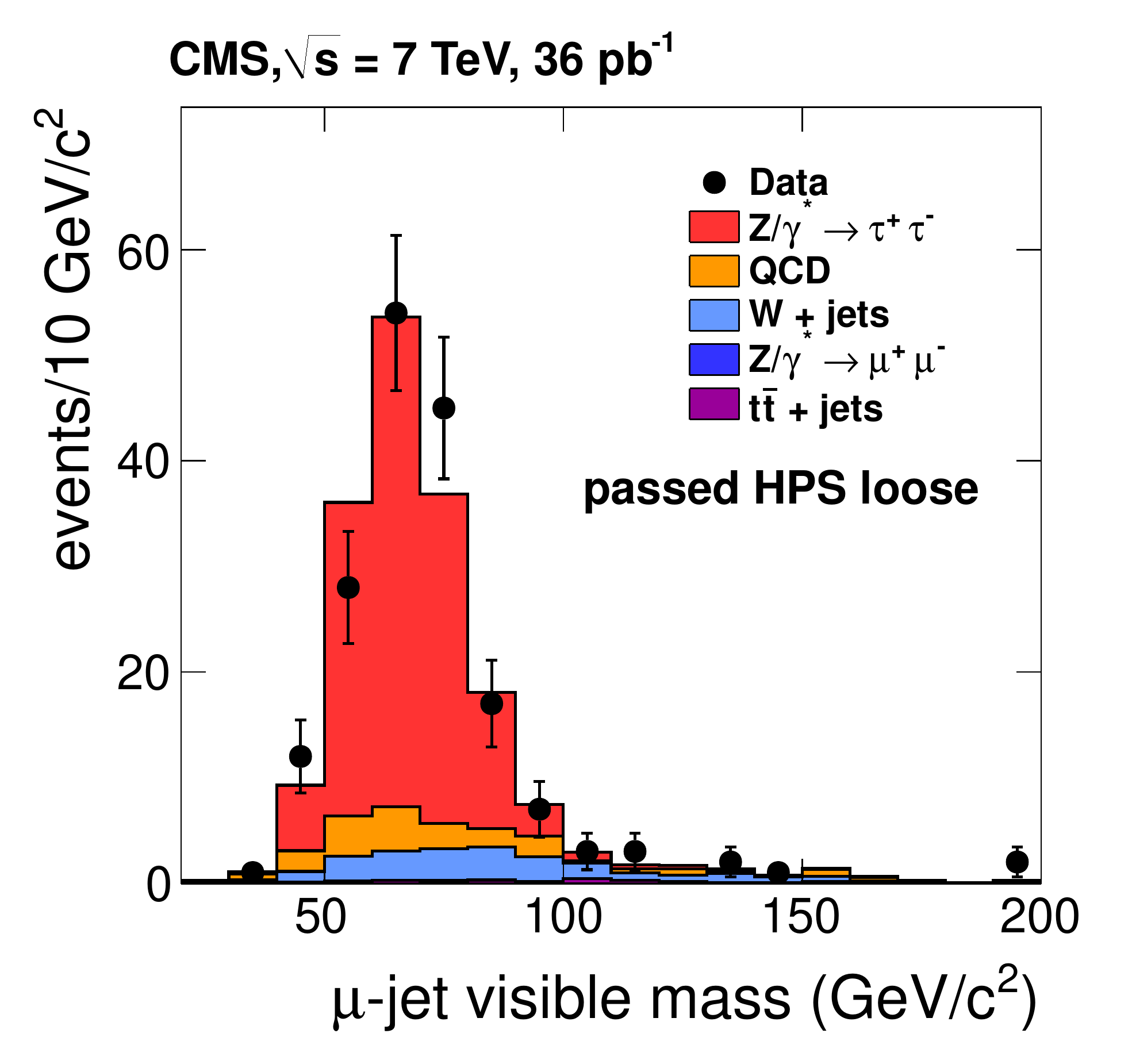}
  \includegraphics[width=0.48\textwidth]{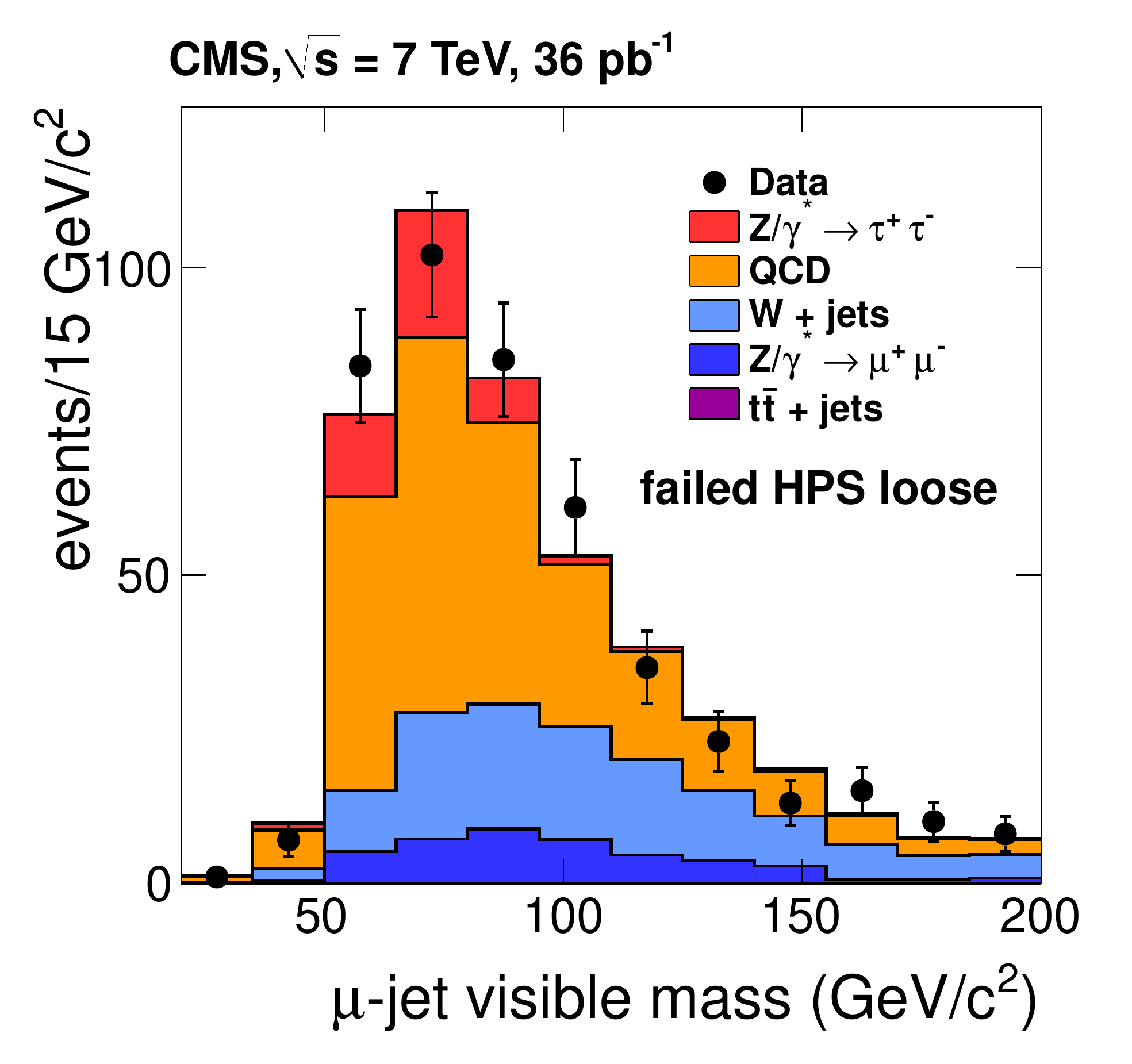}
    \caption{Invariant mass distribution of the $\mu$-jet system
for preselected events  which pass (left) and fail (right) the HPS ``loose'' $\tau_{\rm{h}}$ identification requirements (solid symbols)
             compared to predictions of the MC simulation (histograms).
             }
    \label{fig:efficiency}
  \end{center}
\end{figure}

\begin{table}[htbp]
\begin{center}
\caption{\captiontext
         Efficiency for a $\tau_{\rm{h}}$ to pass the
         HPS and TaNC identification criteria,
         measured by fitting the $\PZ\to\tau\tau$ signal contribution
         in the samples of the ``passed'' and ``failed'' preselected events.
         The uncertainties  of the fit are statistical only.
         The statistical uncertainties of the MC predictions are small
         and can be neglected.
         The last column represents the data-to-MC correction factors and their
         full uncertainties
         including statistical and systematic components.
         Data-to-MC ratios for
         the $\tau_{\rm{h}}$ reconstruction efficiency
         determined using
         fits to the measured $\PZ$ production
         cross sections as described in~\cite{ourZtautaupaper} are also shown.
         }
\tablesize
\begin{tabular}{|l|c|c|c|}
\hline
Algorithm       & Fit data & Expected MC &  Data/MC \\
\hline
HPS  ``loose''  & $0.70 \pm 0.15$ & $0.70$ & $1.00 \pm 0.24$  \\
HPS  ``medium'' & $0.53 \pm 0.13$ & $0.53$ & $1.01 \pm 0.26$ \\
HPS  ``tight''  & $0.33 \pm 0.08$ & $0.36$ & $0.93 \pm 0.25$ \\
\hline
TaNC ``loose''  & $0.76 \pm 0.20$ & $0.72$ & $1.06 \pm 0.30$  \\
TaNC ``medium'' & $0.63 \pm 0.17$ & $0.66$ & $0.96 \pm 0.27$  \\
TaNC ``tight''  & $0.55 \pm 0.15$ & $0.55$ & $1.00 \pm 0.28$ \\
\hline
\hline
HPS  ``loose''  & \multicolumn{2}{c|}{$\tau\tau$ combined fit ~\cite{ourZtautaupaper}} & $0.94 \pm 0.09$ \\
\hline
HPS  ``loose''  & \multicolumn{2}{c|}{$\tau\tau$ to $\mu\mu, ee$ fit ~\cite{ourZtautaupaper}} & $0.96 \pm 0.07$ \\
\hline
\end{tabular}
\label{tab:TauIdEfficiencies}
\end{center}
\end{table}

Results of the fits are summarized in Table~\ref{tab:TauIdEfficiencies}.
The values measured in data, ``Fit data'', are compared
with the expected values, ``Expected MC'', obtained by repeating the
fitting procedure on simulated events.
The efficiency of the $\tau_{\rm{h}}$ algorithms on preselected events is
approximately 30\% higher than for
an inclusive sample, without preselection.
In general
the value of the efficiency depends on the $\pt$ and $\eta$ requirements,
which are applied
in each individual physics analysis. The main goal of
this study is to perform the data-to-MC comparison and to determine
data-to-MC correction factors and their uncertainties.
The agreement in the mean values of the fits between data and MC simulation is
observed to be better than a few percent, although with this data sample,
the statistical uncertainties of the fits are in the range of 20--30\%.

Systematic uncertainties on the measured $\tau_{\rm{h}}$ identification
efficiencies arise from uncertainties on track reconstruction ($4\%$)
and from uncertainties on the probabilities for jets to pass the
``leading'' track $\pt$ and loose isolation requirements applied in
the preselection ($\le 12\%$).  Uncertainties on track momentum and
$\tau_{\rm{h}}$ energy scales have an effect on the measured $\tau_{\rm{h}}$
identification efficiencies below $1\%$. All numbers represent
relative uncertainties.

The resulting ratio of the measured efficiencies to those predicted by
MC simulation for $\tau_{\rm{h}}$ decays to pass the ``loose'', ``medium'',
 and ``tight''
HPS and TaNC working points are presented in the last column of
Table~\ref{tab:TauIdEfficiencies}. The uncertainties on the ratios
represent the full uncertainties of the method, which are calculated by
adding the statistical and systematic uncertainties in quadrature.  The
total uncertainty of the measured efficiency of the $\tau_{\rm{h}}$
algorithms is dominated by the statistical uncertainty of the fit.
The simulation describes the data well.
Since the
same event sample is used to evaluate efficiencies for different
working points, the results are correlated.

The values presented in Table~\ref{tab:TauIdEfficiencies} are used as
inputs for fits to  measure
the uncertainty of the $\tau_{\rm{h}}$
reconstruction and identification efficiency with higher precision
by comparing the yield of the $\PZ \to \tau \tau$
events in different decay modes and the yield of $\PZ\to\mu\mu$ and $\PZ\to e e$
events,
as described elsewhere~\cite{ourZtautaupaper}.
The first approach uses a
simultaneous fit of the four $\PZ \to \tau \tau$ decay channels with
final states $\mu\mu, e\mu, \mu\tau_{\rm{h}}$, and $e \tau_{\rm{h}}$. As a result of
the fit, the combined cross section and $\tau_{\rm{h}}$ efficiency are measured.
The data-to-MC correction factor for the HPS ``loose'' working point
is measured to be $0.94 \pm 0.09$.  The second approach is based on
a comparison of the $\tau_{\rm{h}}$ channels, $\PZ \to \mu\tau_{\rm{h}}$ and $e
\tau_{\rm{h}}$, to the combined $\PZ \to \mu\mu, ee$ cross section as measured by
CMS.  The data-to-MC correction factor for the HPS ``loose'' working
point in this case is measured to be $0.96 \pm 0.07$. The slightly
smaller uncertainty of the latter method is explained by the higher
precision of the combined $\PZ \to \mu\mu, ee$ cross-section
measurement. These values are also presented in
Table~\ref{tab:TauIdEfficiencies}.
Both approaches yield more precise uncertainties, 9\% and 7\%,
than the 24\% from the tag-and-probe method, for the ``loose`` HPS working
point.
To achieve this precision,  the methods  rely on assumptions about the
physics source of the signal, i.e., the values of the
inclusive Z production cross section and $\PZ \to \tau\tau$
branching fraction, and the absence of non-SM sources in the data sample.
In physics analyses where these assumptions cannot be made, such as the
measurement of the $\PZ \to \tau\tau$ production cross section
itself~\cite{ourZtautaupaper}
 and the search for $H \to \tau\tau$~\cite{ourHtautaupaper},
the tag-and-probe method remains the only
 one available.

The expected $\tau_{\rm{h}}$ efficiency values from the
$\PZ \to \tau \tau$ process,
with  a reconstructed
 $\left| \eta_{\tau_{\rm{h}}} \right| < 2.3$,
and either  $\pt^{\tau_{\rm{h}}} > 15$\GeVc or $\pt^{\tau_{\rm{h}}} > 20$\GeVc,
are estimated using simulated events
 and presented in
Table~\ref{tab:EffMC}. The selections are applied both at the generated and reconstructed levels.
A matching of $\Delta R < 0.15$ between the generated and reconstructed $\tau_{\rm{h}}$ directions is required.
Figure~\ref{fig:zttefficiency} shows the expected efficiencies as a function
of the generated
$\pt^{\tau_{\rm{h}}}$ for all working points of each algorithm.

\begin{table}[htbp]
\begin{center}
\caption{\captiontext
         The expected efficiency for $\tau_{\rm{h}}$ decays to pass the
HPS and TaNC identification criteria
         estimated using $\PZ \to \tau \tau$ events from the MC simulation
 for two different selection requirements on
$\pt^{\tau_{\rm{h}}}$. The requirement is applied both at the reconstruction
 and generator levels.
The statistical uncertainties of the MC predictions are smaller
  than the least significant digit of the efficiency values
 in the table and are not shown.}
\tablesize
\begin{tabular}{|l|c|c|c|c|c|c|}
\hline
Algorithm       & \multicolumn{3}{c|}{HPS}
      & \multicolumn{3}{c|}{TaNC}  \\
\hline
       & ``loose'' & ``medium'' & ``tight'' & ``loose'' & ``medium'' & ``tight'' \\
\hline
Efficiency ($\pt^{\tau_{\rm{h}}} > 15$\GeVc) & 0.46  & 0.34 &  0.23
 & 0.54 & 0.43 & 0.30 \\
Efficiency ($\pt^{\tau_{\rm{h}}} > 20$\GeVc) & 0.50 & 0.37 & 0.25
 & 0.58 & 0.48 & 0.36 \\
\hline
\end{tabular}
\label{tab:EffMC}
\end{center}
\end{table}

\begin{figure}[hbtp]
  \begin{center}
  \includegraphics[width=0.48\textwidth]{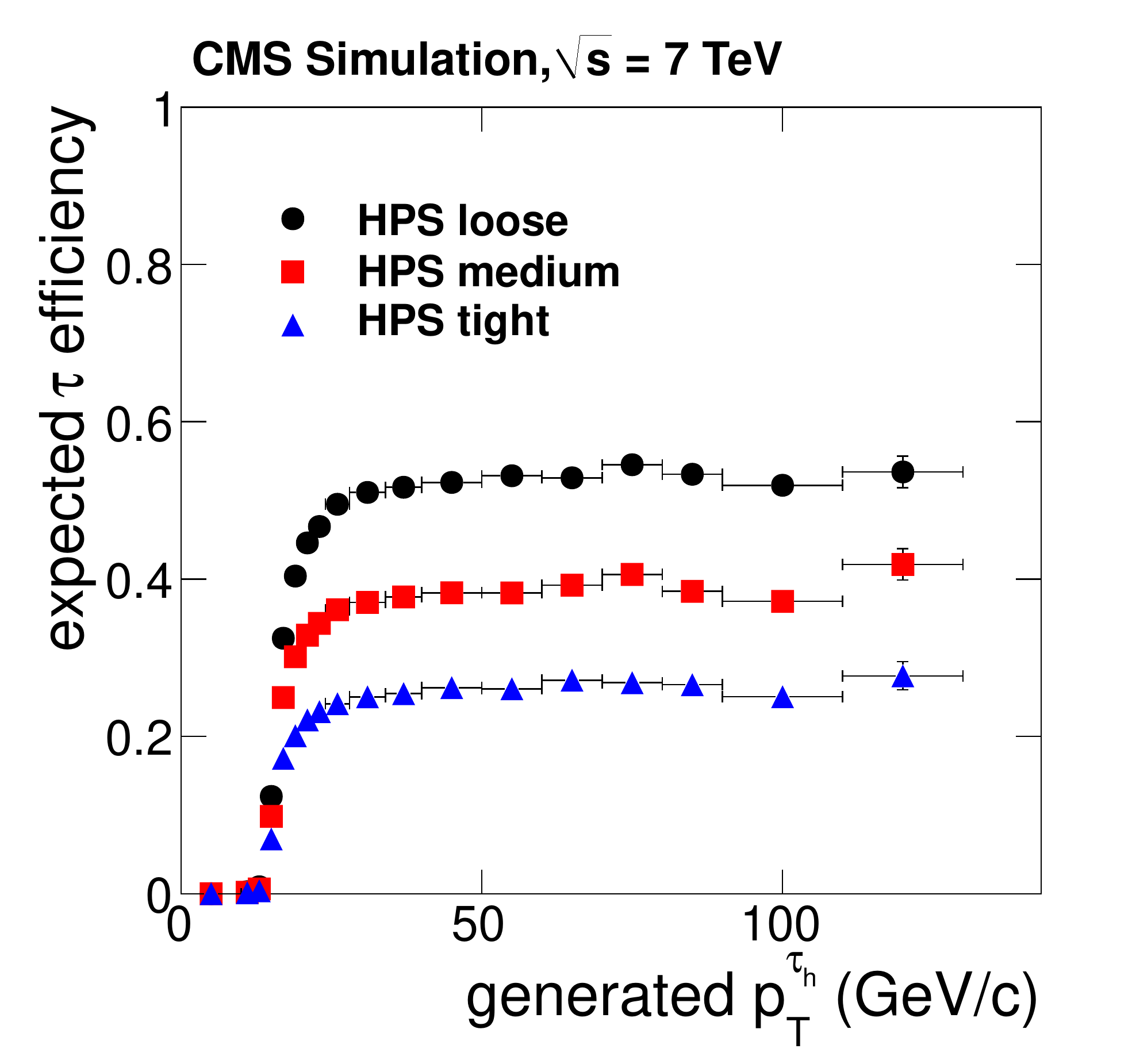}
  \includegraphics[width=0.48\textwidth]{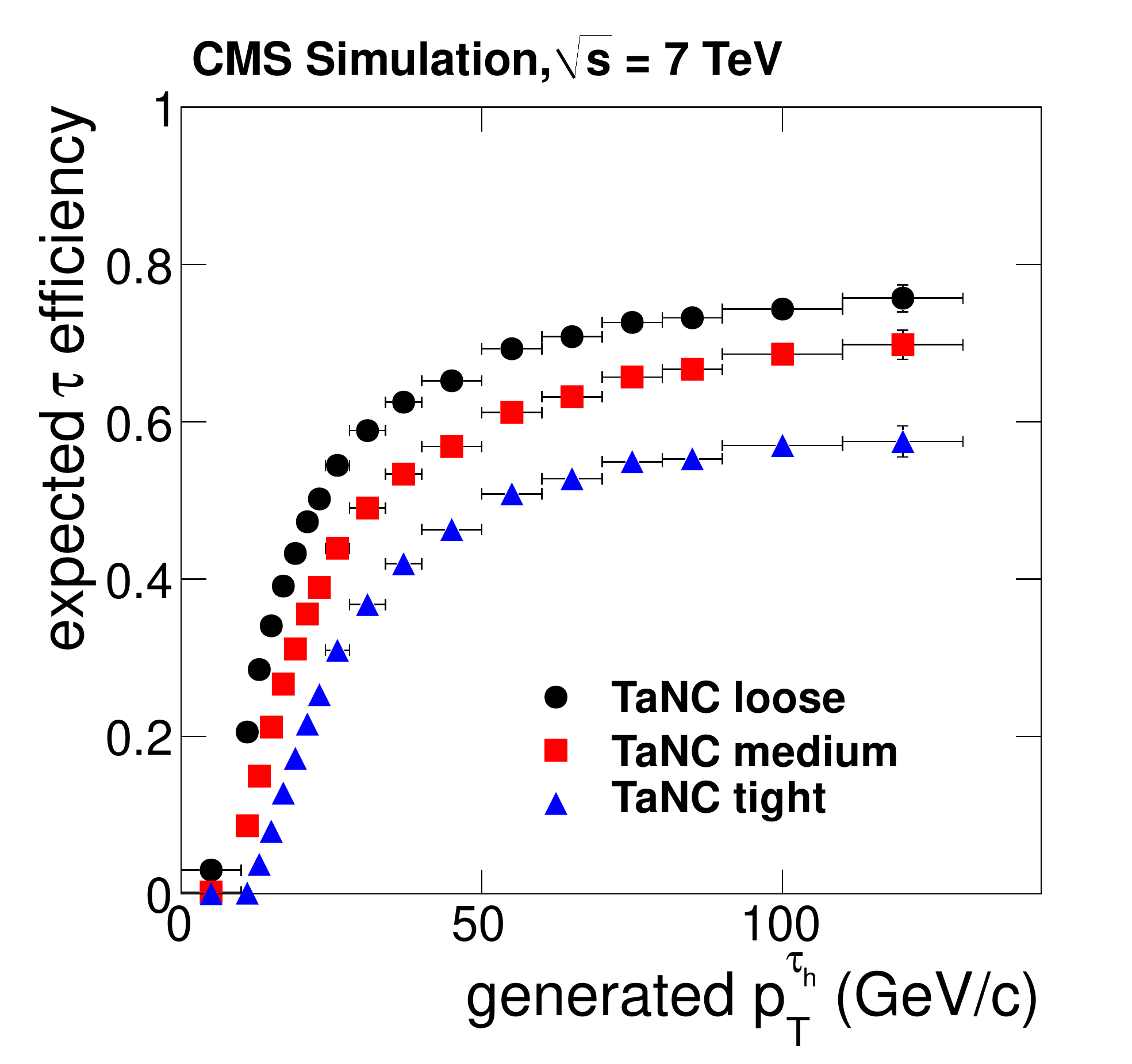}
    \caption{The expected efficiency of the $\tau_{\rm{h}}$ algorithms as a function of generated $\pt^{\tau_{\rm{h}}}$, estimated using
a sample of simulated $\PZ \to \tau \tau$ events for the HPS (left) and TaNC (right) algorithms, for the "loose", "medium", and "tight" working points.
             }
    \label{fig:zttefficiency}
  \end{center}
\end{figure}

\section{Reconstruction of the \texorpdfstring{$\tau_{\rm{h}}$}{tau(h)} Decay Mode}

\begin{figure}[hbtp]
  \begin{center}
\includegraphics[width=0.48\textwidth]{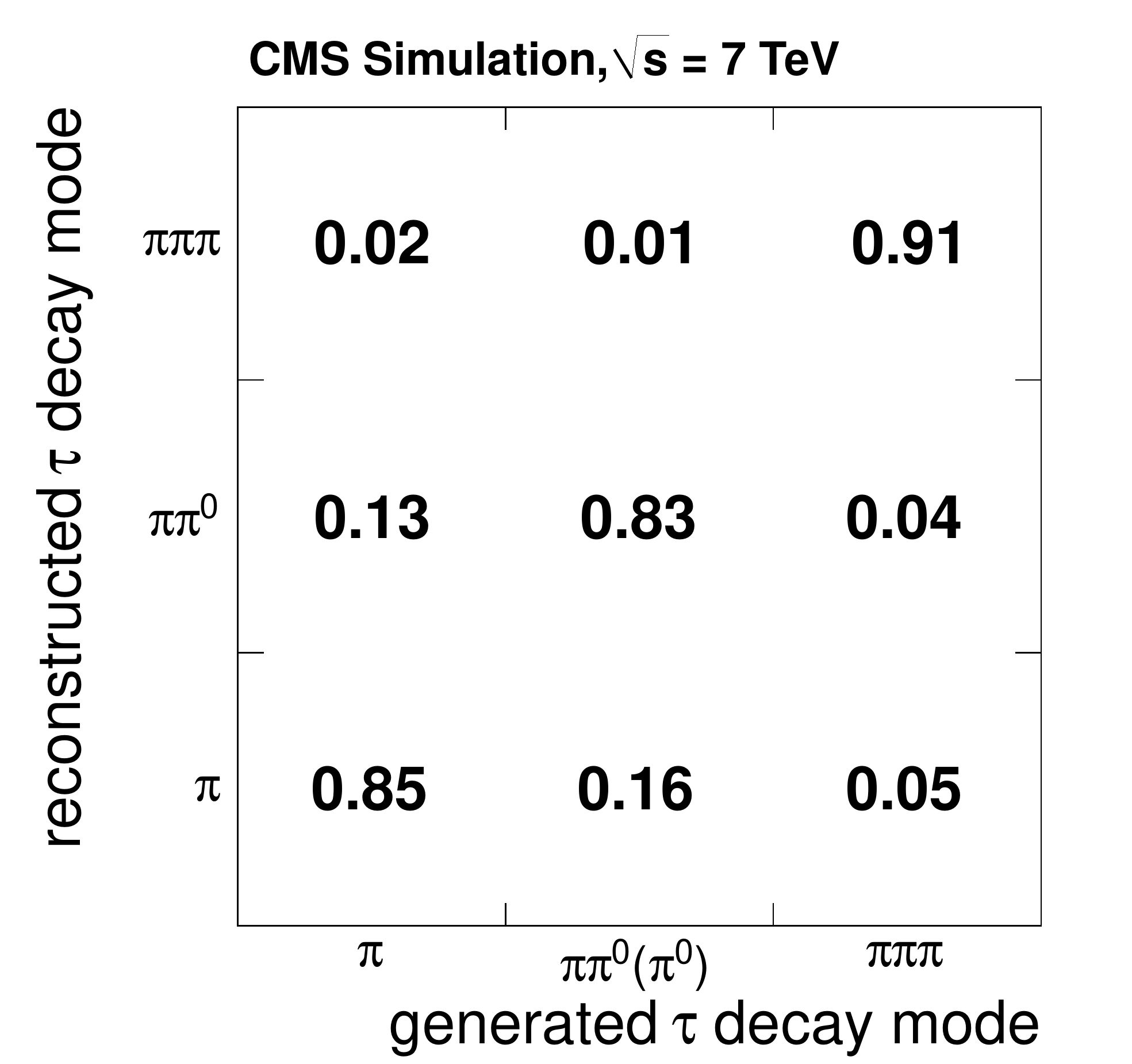}
\includegraphics[width=0.48\textwidth]{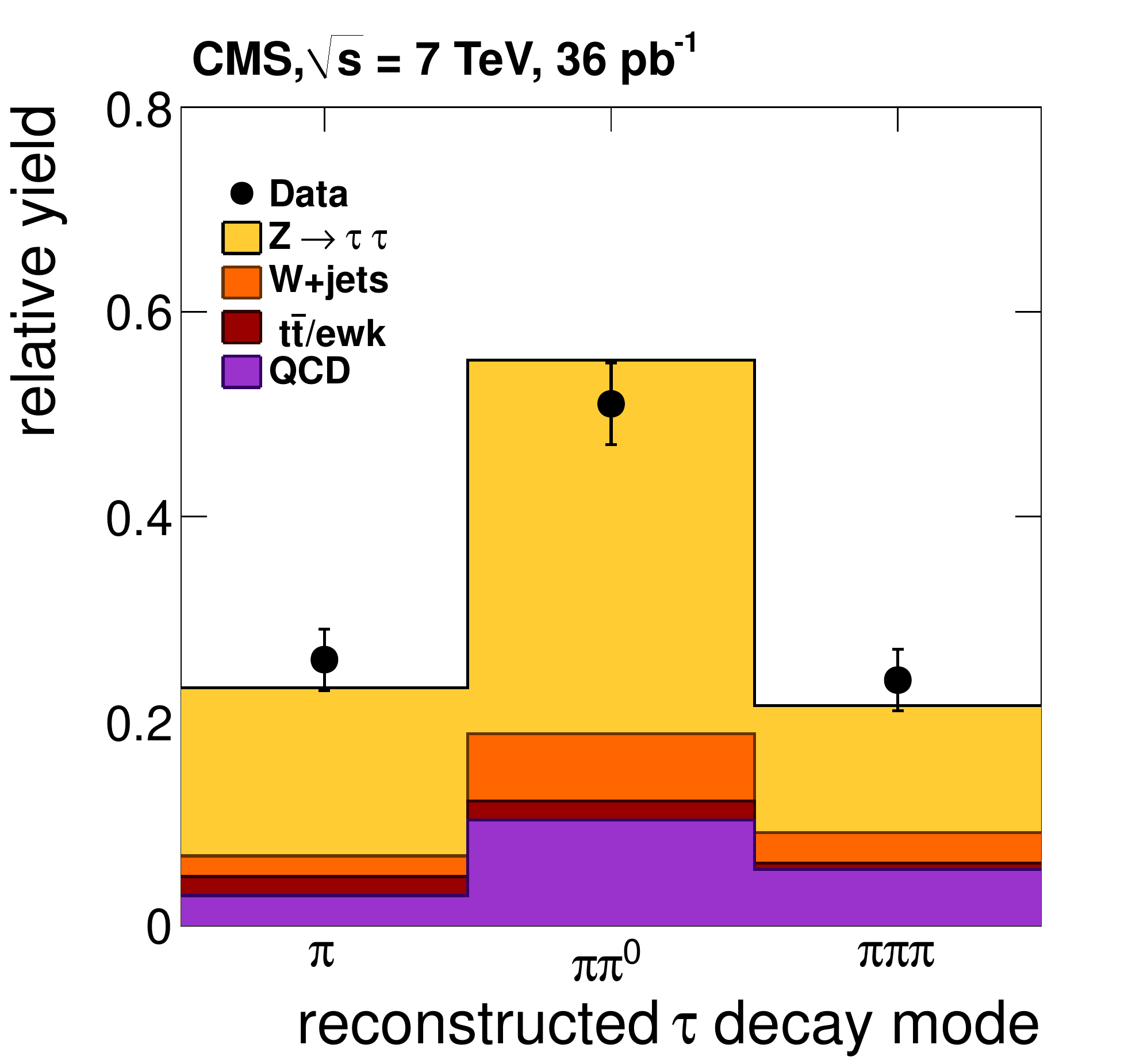}
    \caption{
(left) The fraction of generated $\tau_{\rm{h}}$ decays of a given type
reconstructed in a certain decay mode for the HPS ``loose'' working point
from simulated $\PZ\to\tau\tau$ events.
(right) The relative yield of $\tau_{\rm{h}}$
 reconstructed in different decay modes
in the $\PZ\to\tau\tau\to\mu\tau_{\rm{h}}$ data sample compared to
the MC predictions. The MC simulation is a  mixture of the signal
and background
 samples based on
the corresponding cross sections, as shown by the histograms.}
    \label{fig:decaymodes}
  \end{center}
\end{figure}

The correlation between the generated and reconstructed $\tau_{\rm{h}}$ decay modes
is studied using a sample of simulated $\PZ\to\tau\tau $ events.
The results are presented in Fig.~\ref{fig:decaymodes} (left). Each
column represents one generated decay mode normalized to unity.
Each row corresponds to one reconstructed decay mode.
The numbers
demonstrate the fraction of generated $\tau_{\rm{h}}$ of a
given type
reconstructed in a specific decay mode. Both generated and
reconstructed $\tau_{\rm{h}}$ are required to have a visible transverse
momentum $\pt^{\tau_{\rm{h}}} > 15$\GeVc, and to match within a cone of
$\Delta R = 0.15$.
For each of the generated decay modes, the fraction of
correctly reconstructed decays is more than 80\%, reaching
90\% for the three-charged-pion decay mode.

A data-to-MC comparison of the relative yield of events reconstructed in
different $\tau_{\rm{h}}$ decay modes in a data sample of
$\PZ\to\tau\tau\to\mu\tau_{\rm{h}}$ events
is shown in Fig.~\ref{fig:decaymodes} (right).
The events are selected using the
requirements described
in~\cite{ourZtautaupaper}.
The $\tau_{\rm{h}}$ candidates are required to have visible
transverse momenta $\pt^{\tau_{\rm{h}}} > 20$\GeVc within the geometric
acceptance $\left| \eta \right| < 2.3$.  The MC sample
represents a mixture of the signal and background MC samples based on
the corresponding cross sections. The performance of the $\tau_{\rm{h}}$ algorithm
 is well reproduced by the MC simulation.

\section{Reconstruction of the \texorpdfstring{$\tau_{\rm{h}}$}{tau(h)} Energy}

Since charged hadrons and photons are reconstructed with high
precision using the PF techniques, the reconstructed $\tau_{\rm{h}}$ energy is
expected to be close to the true energy of its visible decay products.
According to simulation, the ratio of the reconstructed to the true
visible $\tau_{\rm{h}}$ energy for the HPS algorithm is constant as a
function of energy  and within
2\% of unity, while for TaNC it decreases by about 2\% as
$\pt^{\tau_{\rm{h}}}$ approaches 60\GeVc.  The $\eta$ dependence is more
pronounced.  For both algorithms the reconstructed $\tau_{\rm{}h}$ energy
is underestimated  by 5\% with respect to the true energy
as one moves towards higher $\eta$ (from
barrel to endcap region).

\begin{figure}[hbtp]
  \begin{center}
\includegraphics[width=0.48\textwidth]{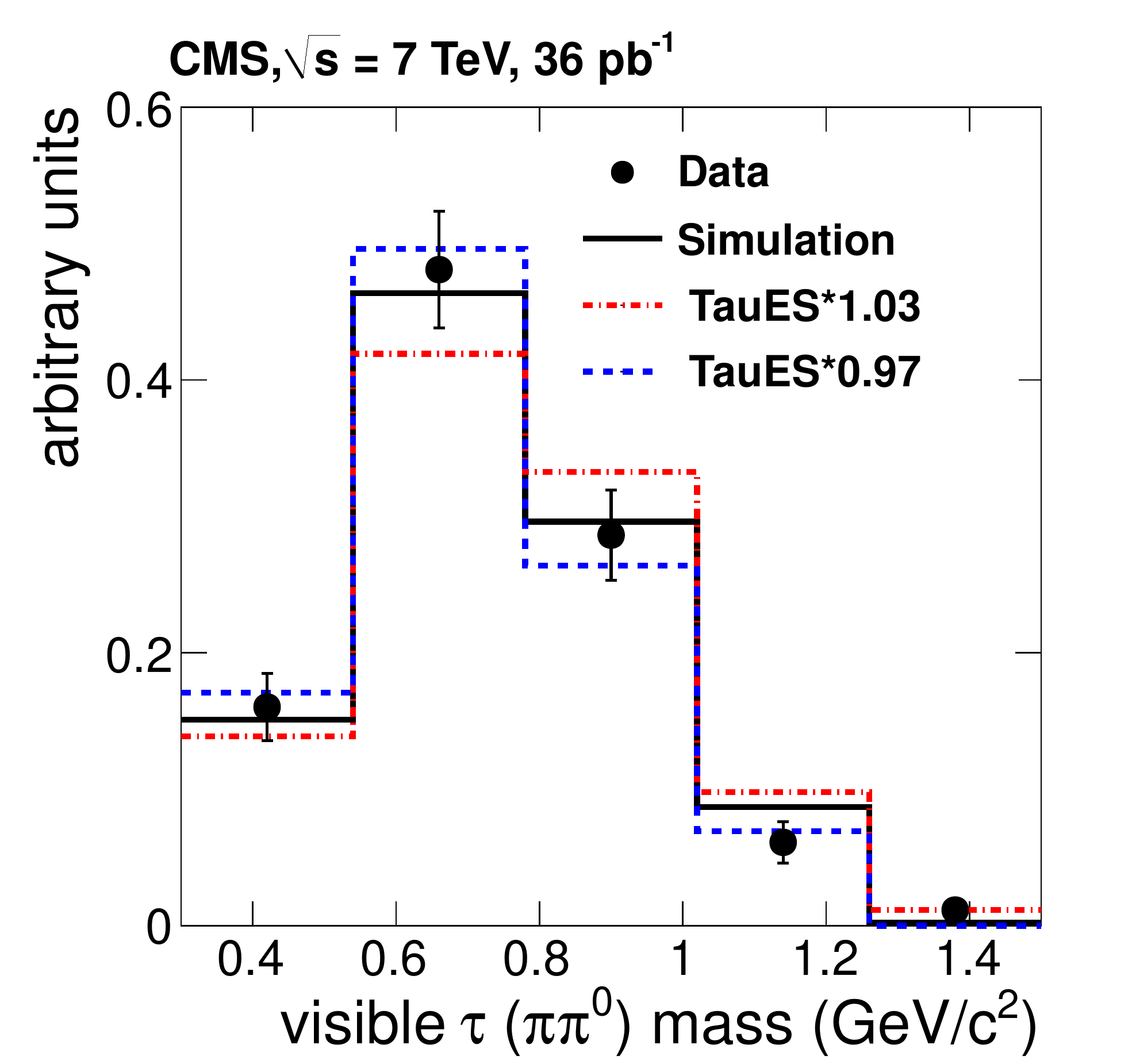}
\includegraphics[width=0.48\textwidth]{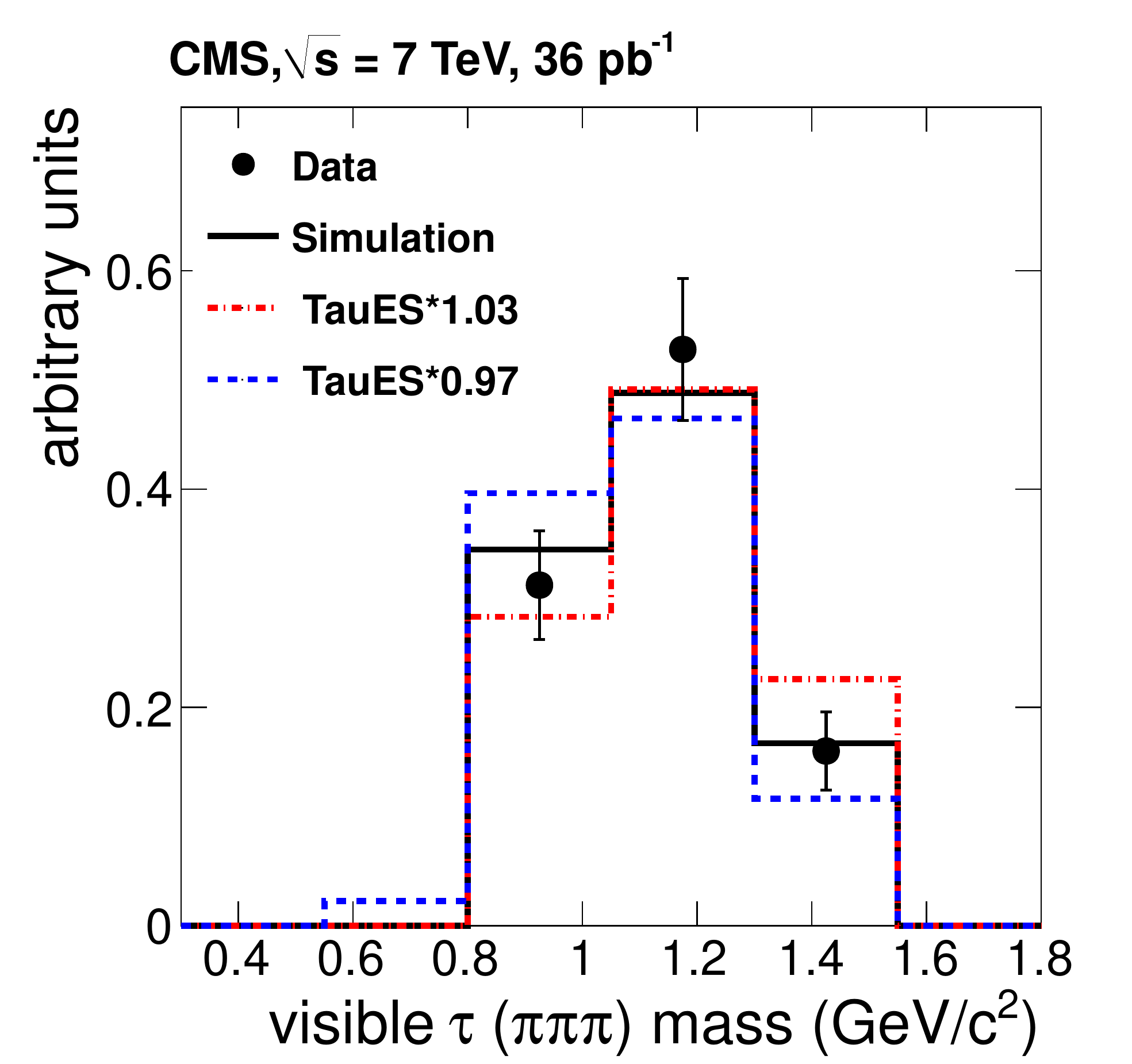}
    \caption{The reconstructed invariant mass  of $\tau_{\rm{h}}$ decaying into one charged and one neutral pion (left) and
into three charged pions (right) from data, compared to predictions of the simulation. The solid lines represent results of the
best fit described in the text and the dashed lines represent the predictions with
the tau energy scale, TauES, varied up and down by 3\% with respect to the best fit value.
             }
    \label{fig:energyscaleuncertainty}
  \end{center}
\end{figure}

The
quality of the $\tau_{\rm{h}}$ energy scale simulation can be examined by
analyzing the $\PZ\to\tau\tau\to\mu\tau_{\rm{h}}$ data sample.
The reconstructed
invariant mass of the $\mu\tau_{\rm{h}}$ system is very sensitive to the
energy scale of the $\tau_{\rm{h}}$, since the muon four-momenta are
measured with high precision. By varying the $\tau_{\rm{h}}$ energy scale
simultaneously in the signal and background MC samples, a set of
templates is produced.  The resulting templates are fitted to the data
and the best agreement is achieved by scaling the $\tau_{\rm{h}}$ energy in
simulation by a factor $0.97 \pm 0.03$, where the uncertainty
is averaged over the pseudorapidity range of the data
sample.

A complementary procedure, which does not assume knowledge of the
$\tau\tau$ invariant mass spectrum, is based on the invariant mass of
reconstructed $\tau_{\rm{h}}$ constituents, shown in
Fig.~\ref{fig:energyscaleuncertainty}. The method uses $\tau_{\rm{h}}$ as
an independent object but relies on good understanding of underlying
background events that contribute to the signal sample.
The fit is performed separately for $\pi \pi^0$ and $\pi \pi \pi$
decay channels, since the major source of the uncertainty is expected
to come from reconstruction of the electromagnetic energy.  The
simulation describes both decay channels well.
The best agreement is achieved by scaling the $\tau_{\rm{h}}$ energy in
simulation by a
factor
$0.97 \pm 0.03$ for the $\pi \pi^0$ decay mode and
by a factor $1.01 \pm 0.02$ for the $\pi \pi \pi$ decay mode.  The effect of the
energy-scale uncertainty on the shape of the $\tau_{\rm{h}}$ invariant mass distribution
 is
also shown in Fig.~~\ref{fig:energyscaleuncertainty}.  Varying
the energy scale in simulation by the uncertainty derived from the
$\mu \tau_{\rm{h}}$ invariant mass fit, i.e. 3\%, corresponds to a significant
deviation in the predicted $\tau_{\rm{h}}$ mass shape.

\section{Measurement of the \texorpdfstring{$\tau_{\rm{h}}$}{tau(h)} Misidentification Rate for Jets}

Jets that could be misidentified as $\tau_{\rm{h}}$  have different
properties depending on their origin.
Most of the jets are produced in QCD processes,
either with or without the associated production
of $\PZ$ or
$\PW$ bosons. To distinguish between them, different data samples are selected.
The QCD-type, gluon-enriched, jets are
selected using events with at least one jet
of transverse momentum $\pt^{\rm{jet}} > 15$\GeVc and a second jet
of $\pt^{\rm{jet}} > 10$\GeVc,
both within $\vert \eta \vert < 2.5$.
The $\PZ$- and $\PW$-type, quark-enriched, jets
are selected by requiring at least
one isolated muon with transverse momentum
$\pt > 15$\GeVc
and  $\vert \eta \vert < 2.1$ and
a jet of transverse momentum $\pt^{\rm{jet}} > 10$\GeVc within $\vert \eta \vert < 2.5$.
In addition, a muon-enriched QCD sample is selected by
requiring a muon and a jet,
but suppressing the $\PW$ contribution by selecting events with $M_{\rm{}T} < 40$\GeVcc.
For each of these samples additional
selection requirements are applied to suppress the background contribution
from events with jets from other sources.

Figure~\ref{fig:fakerate} shows the $\tau_{\rm{h}}$ misidentification rate as a function of the jet $\pt$
for
the ``loose'' working points of the HPS and TaNC algorithms, where
the measured values are compared with the MC predictions for the
different types of jets.
The misidentification rates expected from simulation,
and the measured data-to-MC ratios
are summarized in Table~\ref{tab:fakerate}
for the three working points of both reconstruction
algorithms.
The values are integrated over the $\pt$ and $\eta$ phase space used in the
 $\PZ \to \tau \tau$ analysis, $\pt^{\rm{jet}} > 20$\GeVc and $\vert \eta \vert < 2.3$.
The misidentification rate as a function of reconstruction  efficiency
for all working points of both algorithms is shown in
Fig.~\ref{fig:efffakerate}, which  summarizes the MC estimated
efficiency and the measured misidentification rate values presented in Tables~\ref{tab:EffMC} and
~\ref{tab:fakerate}.
Since the QCD and $\mu$-enriched QCD misidentification rate values
are observed to be similar, only
one set of QCD points is shown.
Open symbols represent results obtained by running an early fixed-cone $\tau_{\rm{h}}$-identification algorithm, used in the
CMS physics technical design report (PTDR,~\cite{PTDR2})
on simulated events. The decay-mode-based HPS and TaNC algorithms perform significantly better than
the fixed-cone algorithm.

\begin{figure}[htbp]
\begin{center}
\includegraphics[width=0.48\textwidth]{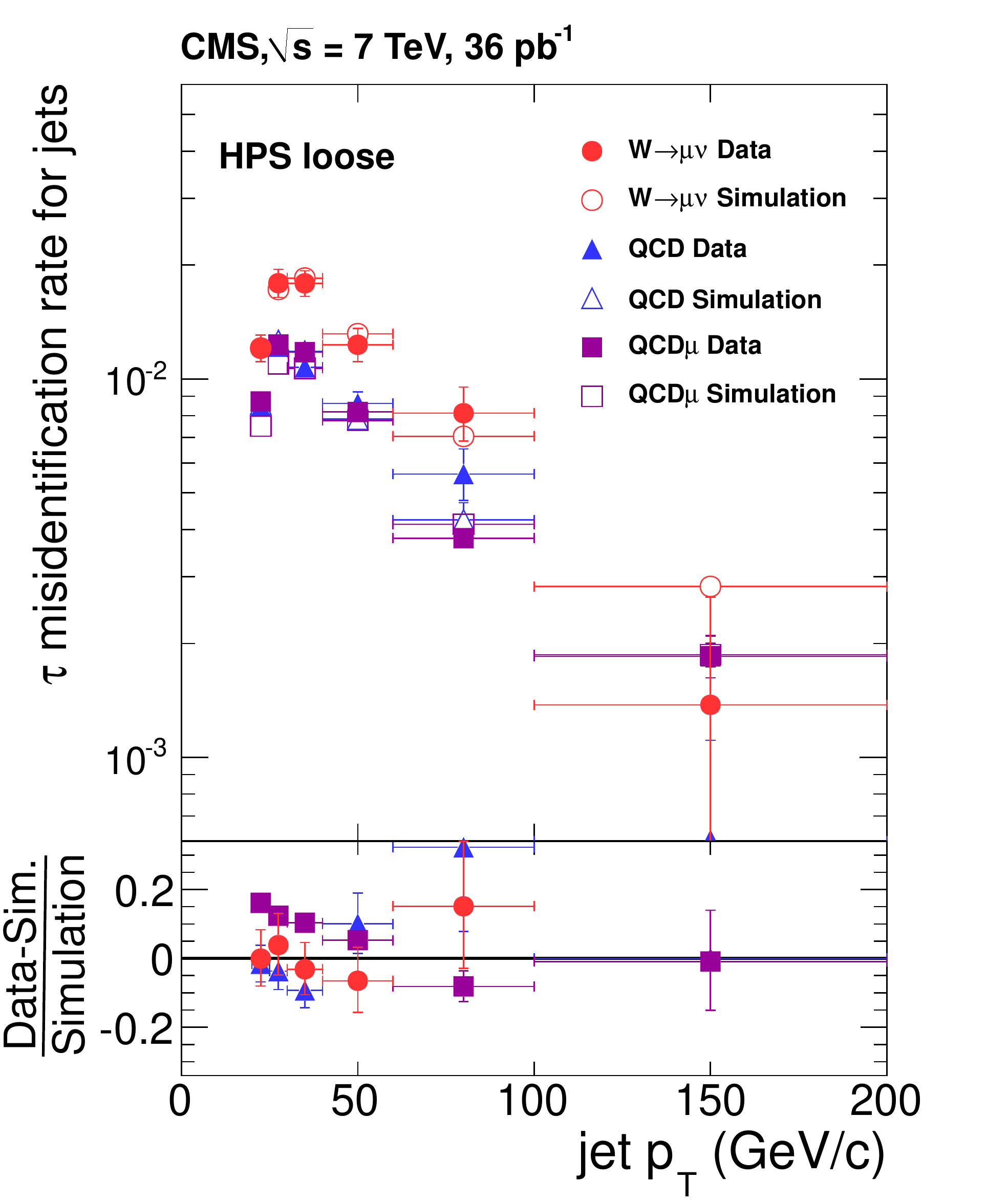}
\includegraphics[width=0.48\textwidth]{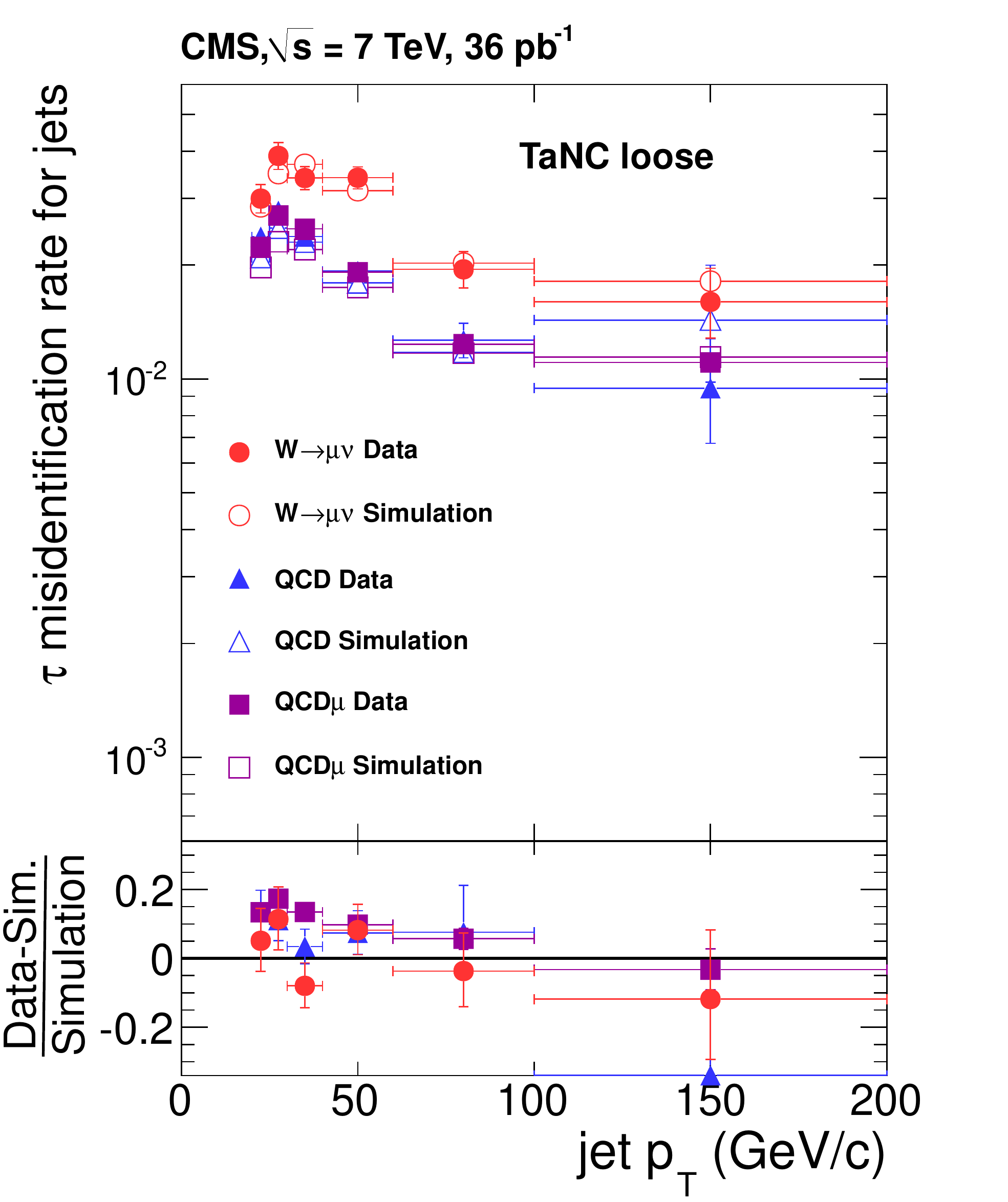}
\caption{Misidentification probabilities for jets to pass
``loose'' working points of the HPS (left) and TaNC (right) algorithms
as a function of jet $\pt$ for QCD, $\mu$-enriched QCD, and $\PW$ type events.
The misidentification rates measured in data are shown by solid symbols and
compared to MC prediction, displayed
with open symbols.}
\label{fig:fakerate}
\end{center}
\end{figure}

\begin{figure}[htbp]
\begin{center}
\includegraphics[width=0.55\textwidth]{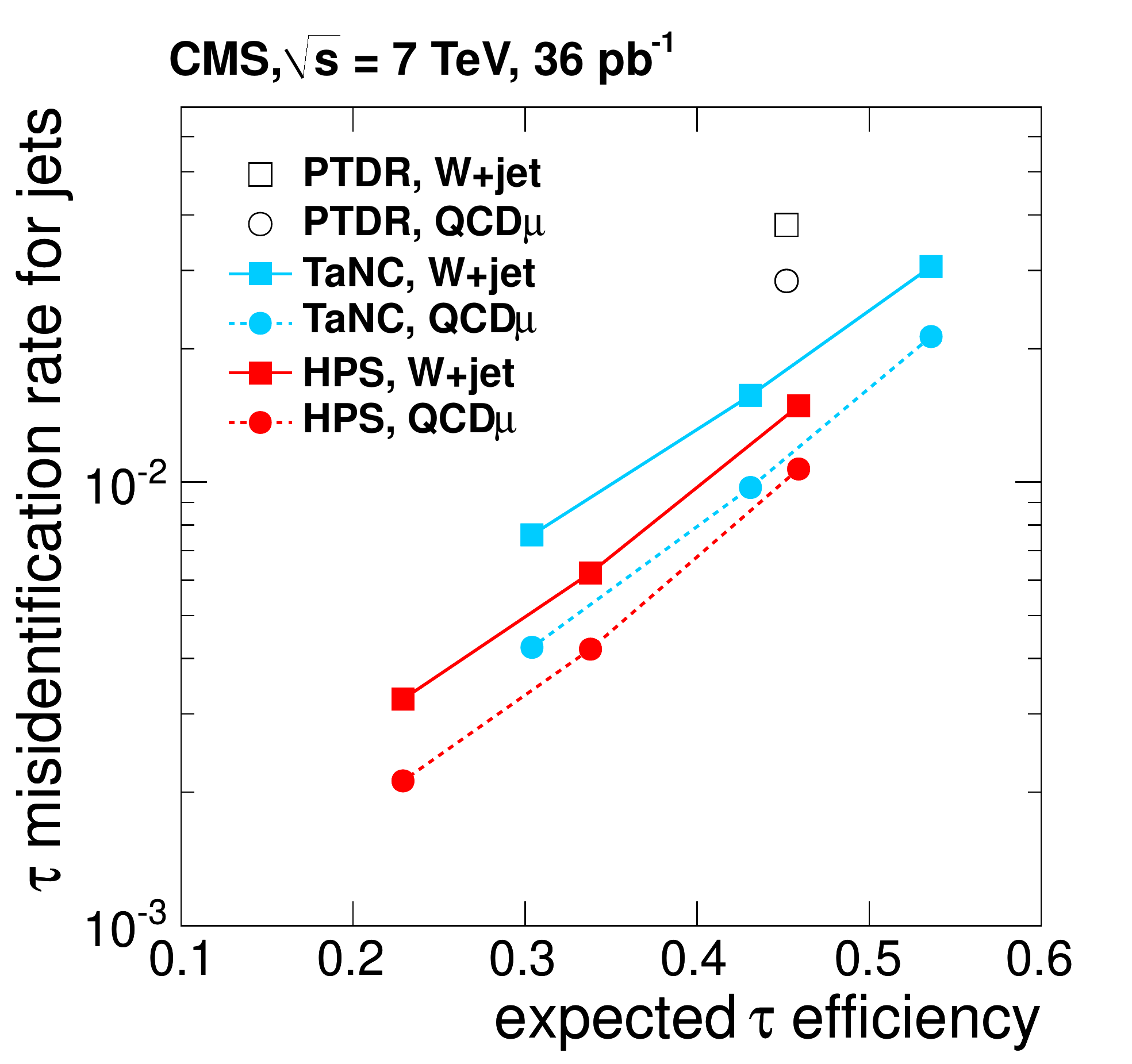}
\caption{
The measured $\tau_{\rm{h}}$ misidentification rate as a function of the
MC-estimated $\tau_{\rm{h}}$ reconstruction
 efficiency for the three working points of the HPS and TaNC algorithms
 from  $\mu$-enriched QCD and $\PW$ data
samples. For each algorithm the ``loose'', ``medium'', and ``tight'' selections
are the points with highest, middle and lowest efficiencies respectively.
The PTDR points represent results of the fixed-cone $\tau_{\rm{h}}$-identification
algorithm~\cite{PTDR2} on simulation.
}
\label{fig:efffakerate}
\end{center}
\end{figure}

\begin{table}[htbp]
\begin{center}
\caption{\captiontext
The MC predicted $\tau_{\rm{h}}$ misidentification rates
and the measured data-to-MC ratios,
 integrated over the $\pt$ and $\eta$ phase space
typical for the $\PZ \to \tau \tau$ analysis.
         }
\tablesize
\begin{tabular}{|l|c|c|c|c|c|c|}
\hline
Algorithm       & \multicolumn{2}{c|}{QCD} & \multicolumn{2}{c|}{QCD$\mu$} & \multicolumn{2}{c|}{$\PW$ + jets} \\
\hline
       & MC (\%) & Data/MC & MC (\%) & Data/MC & MC (\%) & Data/MC \\
\hline
\hline
HPS  ``loose''  & 1.0 & $1.00 \pm 0.04$ & 1.0 & $1.07 \pm 0.01$ & 1.5 & $0.99 \pm 0.04$ \\
HPS  ``medium'' & 0.4 & $1.02 \pm 0.06$ & 0.4 & $1.05 \pm 0.02$ & 0.6 & $1.04 \pm 0.06$ \\
HPS  ``tight''  & 0.2 & $0.94 \pm 0.09$ & 0.2 & $1.06 \pm 0.02$ & 0.3 & $1.08 \pm 0.09$ \\
\hline
TaNC ``loose''  & 2.1 & $1.05 \pm 0.04$ & 1.9 & $1.12 \pm 0.01$ & 3.0 & $1.02 \pm 0.05$ \\
TaNC ``medium'' & 1.3 & $1.05 \pm 0.05$ & 0.9 & $1.08 \pm 0.02$ & 1.6 & $0.98 \pm 0.07$ \\
TaNC ``tight''  & 0.5 & $0.98 \pm 0.07$ & 0.4 & $1.06 \pm 0.02$ & 0.8 & $0.95 \pm 0.09$ \\
\hline
\end{tabular}
\label{tab:fakerate}
\end{center}
\end{table}

\section{Measurement of the \texorpdfstring{$\tau_{\rm{h}}$}{tau(h)} Misidentification Rate for Electrons}

Isolated electrons passing the identification and isolation criteria
of the $\tau_{\rm{h}}$ algorithms are also an important source of background
in many analyses with $\tau_{\rm{h}}$ in the final state.  In this case the
electron is misidentified as a pion originating from $\tau_{\rm{h}}$. A
multivariate discriminant is used to reduce this background, improving
the separation between pions and electrons.
The discriminant is implemented
in the PF algorithm and
its output is
denoted by $\xi$.
The value of the discriminant $\xi$ ranges
between $-1.0$ (most compatible with the pion hypothesis) and $1.0$ (most compatible with the electron hypothesis).

Two selected working points, corresponding to $\xi<-0.1$ and $\xi<0.6$,
are considered in this analysis.
The first working point rejects even those electrons,
that are poorly reconstructed and
is optimized
for a low misidentification rate, about 2\%, at the price of about 4\% losses of
genuine $\tau_{\rm{h}}$.  The second working point suffers from larger
misidentification rates of about 20\%, since it was optimized for $\tau_{\rm{h}}$
efficiencies exceeding 99.5\%.
It rejects only well identified electrons.

The probability for an electron to be misidentified as $\tau_{\rm{h}}$,
 the $e\to\tau_{\rm{h}}$ misidentification rate, is determined using  a sample of isolated electrons
coming from the decay
$\PZ\to ee $.
The events are required to have a reconstructed electron and an
electron that is reconstructed as
$\tau_{\rm{h}}$. The particles must have
opposite charge. The invariant mass of the pair is required to be between
60 and 120\GeVcc.
The tag electron is required to be
isolated and to have a $\pt$ in excess of 25\GeVc.  The second
electron, a probe, is required to pass the HPS ``loose''
working point,
without requiring any specific veto against
electrons, and have
 $\pt$ in excess of 15\GeVc.
The $e\to\tau_{\rm{h}}$ misidentification rate is estimated by measuring the
 ratio between the number of probes passing the
electron-rejection discriminant and the overall number of selected probes.
The sample of events that does not pass the electron-rejection discriminant,
is populated by well-reconstructed electrons. The sample that
passes the discriminant contains poorly reconstructed electrons, as well as
other
background contributions, ``misidentified electrons``.
To remove the
contamination from misidentified electrons, a background subtraction
procedure is performed by fitting the passing and failing $e \tau_{\rm{h}}$ invariant mass
distributions to the superposition of signal and background
components.

Table~\ref{tab:electrons} gives the ratio between the misidentification rates as
measured in the data and those obtained using MC simulation for two $|\eta|$
bins. In the central $\eta$ region, the simulation underestimates the
measured misidentification rates. Within the uncertainties of the
measurement the data-to-MC ratios
 for both discriminants agree in the same $\eta$ intervals.

\begin{table}[htb!]
\centering
\caption{\captiontext
The $e\to\tau_{\rm{h}}$ misidentification rates, found by
applying the tag-and-probe
method to the MC simulation
       and the ratio of the tag-and-probe values obtained in data and MC
simulation, shown in two regions of $\eta$ and for two working points of
the electron-rejection discriminant.
         }
\begin{tabular}{|c|c|c|c|c|}
\hline
Bin & \multicolumn{2}{c|}{Discriminant $\xi<-0.1$} & \multicolumn{2}{c|}{Discriminant $\xi<0.6$}  \\
\hline
$|\eta|$ & MC (\%) & Data/MC & MC (\%) & Data/MC \\
\hline
$<1.5$ & $2.21\pm0.05$ & $1.13\pm0.17$  &  $13.10\pm0.08$ & $1.14\pm0.04$ \\
\hline
$>1.5$ & $3.96\pm0.09$ & $0.82\pm0.18$  &  $26.80\pm0.16$ & $0.90\pm0.04$ \\
\hline
\end{tabular}
\label{tab:electrons}
\end{table}

\section{Summary}

The performances of two reconstruction algorithms for hadronic tau decays developed by
CMS, HPS and TaNC, have been studied using the data sample collected
at a centre-of-mass energy of 7 TeV in 2010 and
corresponding to an integrated luminosity of 36~pb$^{-1}$.  Both
algorithms show good  performance in terms of
$\tau_h$ identification efficiency, approximately 50$\%$, while keeping the
misidentification rate for jets at the level of ${\sim}1\%$.  The MC simulation
was found to describe the data well.  The $\tau_{\rm{h}}$ identification
efficiency was measured with an uncertainty of $24\%$ by using a
tag-and-probe method in  a $\PZ\rightarrow\tau\tau\rightarrow\mu\tau_{\rm{h}}$
data sample, and with an uncertainty of
$7\%$ by using a global fit to all $\PZ\rightarrow \tau\tau$
decay channels and constraining the yield to the measured combined
$\PZ\rightarrow\mu\mu, ee$ cross section.
The scale factor for measured $\tau_{\rm{h}}$
 energies was found to be close to unity with a relative uncertainty less than 3\%.

\section*{Acknowledgments \label{s:ack}}
\hyphenation{Bundes-ministerium Forschungs-gemeinschaft  Forschungs-zentren}
We wish to congratulate our colleagues in the CERN accelerator departments for the excellent performance of
the LHC machine.
We thank the technical and administrative staff at CERN and other CMS
institutes.  This work was supported by
the Austrian Federal Ministry of Science and Research;
the Belgium Fonds de la Recherche Scientifique, and Fonds voor Wetenschappelijk Onderzoek;
the Brazilian Funding Agencies (CNPq, CAPES, FAPERJ, and FAPESP);
the Bulgarian Ministry of Education and Science;
CERN;
the Chinese Academy of Sciences, Ministry of Science and Technology, and National Natural Science Foundation of China;
the Colombian Funding Agency (COLCIENCIAS);
the Croatian Ministry of Science, Education and Sport;
the Research Promotion Foundation, Cyprus;
the Estonian Academy of Sciences and NICPB;
the Academy of Finland, Finnish Ministry of Education and Culture, and Helsinki Institute of Physics;
the Institut National de Physique Nucl\'eaire et de Physique des
    Particules~/~CNRS, and Commissariat \`a l'\'Energie Atomique et aux \'Energies Alternatives~/~CEA, France;
the Bundesministerium f\"ur Bildung und Forschung,
Deutsche Forschungsgemeinschaft, and
Helmholtz-Gemeinschaft Deutscher Forschungszentren, Germany;
the General Secretariat for Research and Technology, Greece;
the National Scientific Research Foundation, and National Office for Research and Technology, Hungary;
the Department of Atomic Energy and the Department of Science and Technology, India;
the Institute for Studies in Theoretical Physics and Mathematics, Iran;
the Science Foundation, Ireland;
the Istituto Nazionale di Fisica Nucleare, Italy;
the Korean Ministry of Education, Science and Technology and the World Class University program of NRF, Korea;
the Lithuanian Academy of Sciences;
the Mexican Funding Agencies (CINVESTAV, CONACYT, SEP, and UASLP-FAI);
the Ministry of Science and Innovation, New Zealand;
the Pakistan Atomic Energy Commission;
the State Commission for Scientific Research, Poland;
the Funda\c{c}\~ao para a Ci\^encia e a Tecnologia, Portugal;
JINR (Armenia, Belarus, Georgia, Ukraine, Uzbekistan);
the Ministry of Science and Technologies of the Russian Federation, the Russian Ministry of Atomic Energy and the Russian Foundation for Basic Research;
the Ministry of Science and Technological Development of Serbia;
the Ministerio de Ciencia e Innovaci\'on, and Programa Consolider-Ingenio 2010, Spain;
the Swiss Funding Agencies (ETH Board, ETH Zurich, PSI, SNF, UniZH, Canton Zurich, and SER);
the National Science Council, Taipei;
the Scientific and Technical Research Council of Turkey, and Turkish Atomic Energy Authority;
the Science and Technology Facilities Council, UK;
the US Department of Energy, and the US National Science Foundation.

Individuals have received support from
the Marie-Curie programme and the European Research Council (European Union);
the Leventis Foundation;
the A. P. Sloan Foundation;
the Alexander von Humboldt Foundation;
the Belgian Federal Science Policy Office;
the Fonds pour la Formation \`a la Recherche dans l'Industrie et dans l'Agriculture (FRIA-Belgium);
the Agentschap voor Innovatie door Wetenschap en Technologie (IWT-Belgium);
and the Council of Science and Industrial Research, India.

\fi

\ifx\ver\verAN
\input{introduction}

\input{taualgorithms}

\input{tauid_usingtau}

\input{tauid_usingmumu}

\input{tauid_usingmc}

\input{tau_energyscale}

\input{tau_energyscaleuncertainty}

\input{tau_electronsdiscrimination}

\input{fakerate_summary}
\fi

\bibliography{auto_generated}   

\cleardoublepage \appendix\section{The CMS Collaboration \label{app:collab}}\begin{sloppypar}\hyphenpenalty=5000\widowpenalty=500\clubpenalty=5000\textbf{Yerevan Physics Institute,  Yerevan,  Armenia}\\*[0pt]
S.~Chatrchyan, V.~Khachatryan, A.M.~Sirunyan, A.~Tumasyan
\vskip\cmsinstskip
\textbf{Institut f\"{u}r Hochenergiephysik der OeAW,  Wien,  Austria}\\*[0pt]
W.~Adam, T.~Bergauer, M.~Dragicevic, J.~Er\"{o}, C.~Fabjan, M.~Friedl, R.~Fr\"{u}hwirth, V.M.~Ghete, J.~Hammer\cmsAuthorMark{1}, S.~H\"{a}nsel, M.~Hoch, N.~H\"{o}rmann, J.~Hrubec, M.~Jeitler, W.~Kiesenhofer, M.~Krammer, D.~Liko, I.~Mikulec, M.~Pernicka, B.~Rahbaran, H.~Rohringer, R.~Sch\"{o}fbeck, J.~Strauss, A.~Taurok, F.~Teischinger, C.~Trauner, P.~Wagner, W.~Waltenberger, G.~Walzel, E.~Widl, C.-E.~Wulz
% ...
\vskip\cmsinstskip
% ...
1:~~Also at CERN, European Organization for Nuclear Research, Geneva, Switzerland\\
% ...

\end{sloppypar}
\end{document}